\documentclass[twocolumn,showpacs,preprintnumbers,amsmath,amssymb,prl,superscriptaddress]{revtex4}
\usepackage{graphicx}

\begin{document}

\title{\bf Correlated adatom trimer on metal surface:\\ A
continuous time quantum Monte Carlo study}

\author{V. V. Savkin}
\email{savkin@sci.kun.nl} \affiliation{Institute of Molecules and
Materials, University of Nijmegen, 6525 ED Nijmegen, The
Netherlands}
\author{A. N. Rubtsov}
\affiliation{Department of Physics, Moscow State University,
119992 Moscow, Russia }
\author{M. I. Katsnelson}
\affiliation{Institute of Molecules and Materials, University of
Nijmegen, 6525 ED Nijmegen, The Netherlands}
\author{A. I. Lichtenstein}
\affiliation{Institute of Theoretical Physics, University of
Hamburg, 20355 Hamburg, Germany} 

\begin{abstract}
The problem of three interacting Kondo impurities is solved within
a numerically exact continuous time quantum Monte Carlo scheme. A
suppression of the Kondo resonance by interatomic exchange
interactions for different cluster geometries is investigated. It is
shown that a drastic difference between the Heisenberg and Ising
cases appears for antiferromagnetically coupled adatoms. The
effects of magnetic frustrations in the adatom trimer are
investigated, and possible connections with available experimental
data are discussed.
\end{abstract}

\maketitle

The electronic structure of adatoms and clusters on surfaces
constitutes one of the most fascinating subjects in condensed
matter physics and modern nanotechnology \cite{plummer}. The
scanning tunneling microscopy (STM) or spectroscopy technique
allows the study of atomic structure, the electronic energy
spectrum, and magnetic properties of different surfaces at an
atomic scale. In particular, STM gives the unique opportunity of
directly investigating an essentially many-body phenomenon, namely
the Kondo effect \cite{STM,coral,kol}. Earlier only indirect
methods such as analysis of temperature and magnetic field
dependencies of thermodynamic and transport properties were
available \cite{gruner,hewson}. Recently STM studies of small
transition metal nanoclusters on different surfaces have been
performed, including Co dimers \cite{CoAu} and Cr trimers
\cite{CrAu} on a Au surface, and Co clusters on carbon nanotubes
\cite{nanotube}. The electron spectrum of these nanosystems, in
particular the existence of the Kondo resonance, turns out to be
very sensitive to the geometry of the clusters as well as to the
type of magnetic adatoms. The later can be important for
nanotechnological fine tuning of surface electronic structure.

The ``quantum-corral'' type of STM-experiments provides an unique
opportunity to investigate in detail an interplay between the
single-impurity Kondo effect and interatomic magnetic interactions
in nanoclusters which is a key phenomenon in Kondo lattice physics
\cite{doniach,irkhin,si,castro,zhu}. The interaction between
itinerant electrons and localized ones leads to the screening of
the magnetic impurity moment which is the Kondo effect; on the
other hand, the RKKY exchange interaction between localized spins
suppresses the Kondo resonance at the Fermi level. As a result, a
very complicated phase diagram can be obtained with regions
described by a strong coupling regime, ``normal'' magnetic
behavior with logarithmic corrections, and non-Fermi-liquid
behavior \cite{irkhin}. Quantum critical points at the boundary of
different phases is a subject of special interest \cite{si,zhu}.
There exists a general belief that anomalous features of many
$f$-electron systems such as heavy-fermion or non-Fermi-liquid
behavior can be treated in terms of the Kondo lattice picture
\cite{hewson,stewart,amato}. This is why non-perturbative
investigations of the basic physical features of few-atom magnetic
clusters in a metallic medium is of primary interest. At the same
time, due to the extreme complexity of the problem, theoretical
investigations of electronic structure for several Kondo centers
usually involve some uncontrollable approximations, such as a
replacement of the Heisenberg interatomic exchange interactions by
the Ising ones \cite{zhu} or a variational approach based on a
simple trial function \cite{uzdin}.

In this Letter we present results of a numerically exact solution
of the three Kondo impurity problem within the recently developed
continuous time quantum Monte Carlo (CT-QMC) method
\cite{rubtsov}. For the antiferromagnetic (AFM) exchange
interatomic interaction, in contrast to the ferromagnetic (FM)
one, the results for the Heisenberg and Ising systems differ
essentially. Based on our theoretical analysis, the recent
paradoxical experimental results \cite{CrAu} where the Kondo
resonance is observable for an isosceles magnetic triangle but not
for the perfect Cr-trimer or individual Cr adatom will be
discussed.

We start with the system of three impurity correlated sites with
Hubbard repulsion $U$ in a metallic bath and with an effective
exchange interaction $J_{ij}$ between them, a minimal model which
however includes all relevant interactions necessary to describe
magnetic nanoclusters on a metallic surface. The effective action
for such cluster in a metallic medium has the following form:
\begin{eqnarray}\label{S}
&S=S_0+W, \\
&S_0=-\int _0^{\beta} \int _0^{\beta} d\tau d\tau'
\sum_{i,j;\sigma} c^\dag_{i\sigma}(\tau) {\cal{G}}
^{-1}_{ij}(\tau-\tau')c_{j\sigma}(\tau'), \nonumber \\
&W=\int _0^{\beta} d\tau \left( U\sum_{i}n_{i\uparrow}(\tau)
n_{i\downarrow}(\tau)+\sum_{i,j} J_{ij} \mathbf{S}_i(\tau)
\mathbf{S}_j(\tau)\right). \nonumber
\end{eqnarray}
The last term in the right-hand-side of Eq.(\ref{S}) allows us to
consider the most important ``Kondo lattice'' feature, that is,
the mutual suppression of the Kondo screening and intersite
exchange interactions \cite{doniach,irkhin}. Another factor, the
coherence of the resonant Kondo scattering, is taken into account
by the introduction of inter-impurity hopping terms $t_{ij}$ to
the bath Green function which is supposed to be
${\cal{G}}^{-1}_{ij}={\cal{G}}^{-1}_{i}\delta_{ij}-t_{ij}$. Here
${\cal{G}}^{-1}_{i}(i\omega_n)=\mu + i (\omega_n +
\sqrt{\omega_n^2+1})/2$ corresponds to the semicircular density of
states (DOS) with band-width 2 and $t_{ij}$ are inter-impurity
hopping integrals. For real adatom clusters the exchange
interactions are mediated by conduction electrons (RKKY
interactions) which are dependent on the specific electronic
structure of both adatoms and host metal. To simulate this effect
we will consider $J_{ij}$ as independent parameters which is a
common practice in the Kondo lattice problem
\cite{doniach,irkhin,zhu}; otherwise for the half-filled
non-degenerate Hubbard model used in our calculations the exchange
is always antiferromagnetic. In the model (\ref{S}) the geometry
of the problem is specified by the values of exchange integrals
$J_{ij}$ and hopping parameters $t_{ij}$\cite{georges}. We will
concentrate on the case of equilateral triangle when $J_{ij}=J$
and $t_{ij}=t$; to compare with the experimental situation in
Ref.\onlinecite{CrAu} also an isosceles triangle will be
considered. To check an approximation used in Ref.\onlinecite{zhu}
we will investigate the case when all spin-flip exchange terms are
ignored and the Heisenberg ($\mathbf{S}\mathbf{S}$) form of
interaction $J_{ij} \mathbf{S}_i(\tau) \mathbf{S}_j(\tau)$ is
transformed into the Ising ($S_zS_z$) one $J_{ij} S^z_i(\tau)
S^z_j(\tau)$.

We use the numerically exact CT-QMC method \cite{rubtsov} for our
computer simulations. Unlike the Hirsh-Fye discrete-time scheme
\cite{hirsch} it does not involve auxiliary Ising spins, but
performs a random walk in the space of terms of the perturbation
expansion for the Green function. One of the advantages of this
novel approach is the opportunity it provides to study systems
with non-local (in space and in time) interactions, which in the
usual Hirsh-Fye approach would involve a huge increase in the
required number of auxiliary fields and time slices
\cite{rubtsov}.

In brief, numerical simulations make use of the division of the
action into to Gaussian and interaction parts. The later can be
presented in the form $\int d \tau \sum w_{iji'j'}(\tau)$, where
$w=u_{iji'j'} (c^{\dag}_{i, \uparrow} c_{j, \uparrow} -
\alpha_{\uparrow}) (c^{\dag}_{i', \downarrow} c_{j', \downarrow} -
\alpha_{\downarrow})$, where the $\alpha$'s are $c$-numbers chosen
in a special way to minimize the sign problem in the QMC
simulation. It is worth noting that only the presence of these
$\alpha's$ makes possible the simulation of systems with repulsive
forces \cite{rubtsov}.

The formal interaction-representation expansion in powers of
the interaction for the Green function $G_{ij}(\tau,\tau')$ reads
\begin{equation}\label{G}
\sum_k \sum_{I_1, ... , I_k} \int d\tau_1 ...  d\tau_k \frac{{\rm
Tr} \left(T c^\dag_{i \tau'} c_{j \tau} w_{I_1}(\tau_1) ...
w_{I_k}(\tau_k) \right)}{Z k!}.
\end{equation}
Here $Z$ is the partition function and $I$ denotes the set $i, j,
i', j'$. Our algorithm performs the random-walk in a space of all
possible values for $k; \tau_1 ... \tau_k; I_1 ... I_k$. The
numerical averaging of Eq.(\ref{G}) over this random walk gives
the desired Green function for the interacting system with the
action (\ref{S}). Typical value for $k$ in our calculation is
$k\approx \int d\tau \sum ||w_I(\tau)||\approx N \beta U+ N^2
\beta J$, where $\beta$ is the inverse temperature and $N$ is
number of atoms in the cluster. One can see that for the case $U
\gg N J$ the exchange interaction indeed does not slow the
calculation down, its complexity is determined by the local
(Coulomb) interaction.

In order to check the CT-QMC algorithm for a system with
complicated Heisenberg interactions we apply this method to a
simple Hamiltonian analogue of the model (\ref{S}), {\it i.e.} for
${\cal{G}}^{-1}_{i}(i\omega_n)=\mu + i \omega_n$. We compare our
CT-QMC approach with the solution obtained using the exact
diagonalization method. Results for the system with AFM $J$
($J>0$) are shown in Fig.1 for $S_zS_z$ and $\mathbf{S}\mathbf{S}$
interactions. The estimated errorbar in numerical data is
$10^{-3}$ or less.

Although the problem of DOS-calculations from $G(\tau)$ data is
ill-defined, quite reliable estimations can be made at our level
of the numerical accuracy. For example, DOS of the above-mentioned
Hamiltonian models is at most contributed by the four
$\delta$-peaks located at $\pm U/2, \pm (U/2+J)$ for Ising and
$\pm U/2, \pm (U/2+2 J)$ for Heisenberg interaction. Fit of the
numerical data with several $\delta$-peaks indeed resolves their
positions (with a $5\%$ errorbar) and relative heights (with a
$20\%$ errorbar). Further, we study models with a continuous DOS,
and standard maximum-entropy analytical continuation method
\cite{maxent} is used to recover DOS. Normally it resolves the DOS
features from imaginary time QMC data with a similar accuracy.

\begin{figure}
\includegraphics[width=\columnwidth]{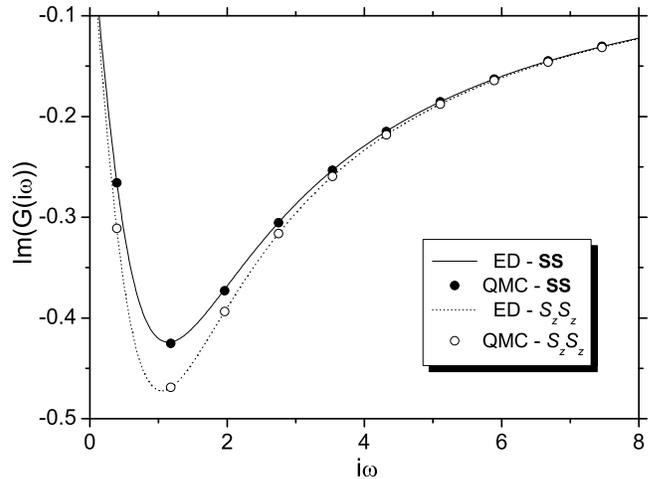}
\caption{\label{fig:epsart} Imaginary part of the Green function
at Matsubara frequencies for the model (\ref{S}) with
${\cal{G}}^{-1}_{i}(i\omega_n)=\mu + i \omega_n$ for
$\mathbf{S}\mathbf{S}$ and $S_zS_z$ interactions. Symbols are
CT-QMC data, lines are exact diagonalization (ED) results.
Parameters of the model: $U=2, J=0.2, t=0, \beta=8, \mu=U/2$. }
\end{figure}

Let us discuss correlated adatom trimer in the metallic bath
depending on type of the effective exchange interaction ($S_zS_z$
or $\mathbf{S}\mathbf{S}$) for AFM and FM cases. First we show
that $\mathbf{S}\mathbf{S}$ type of interaction suppresses the
resonance in AFM case. We study the equilateral triangle at
half-filling. The Green functions at the Matsubara frequencies
obtained by the CT-QMC technique and corresponding DOS are
presented in Fig. 2. The case of $J=0$ corresponds to the single
Kondo impurity in the bath. One can see that there is no essential
difference between $S_zS_z$ and $\mathbf{S}\mathbf{S}$ types of
interaction in FM but for the AFM one this difference is very
important. The AFM $\mathbf{S}\mathbf{S}$ interaction leads to
pronounced suppression of the Kondo resonance at Fermi level for
physically relevant values of $J$ (Fig. 2). On the other hand, in
the FM case we do not observe any essential difference between the
$S_zS_z$ and $\mathbf{S}\mathbf{S}$ types of the interaction for a
wide range of the model parameters.

\begin{figure}
\includegraphics[width=\columnwidth]{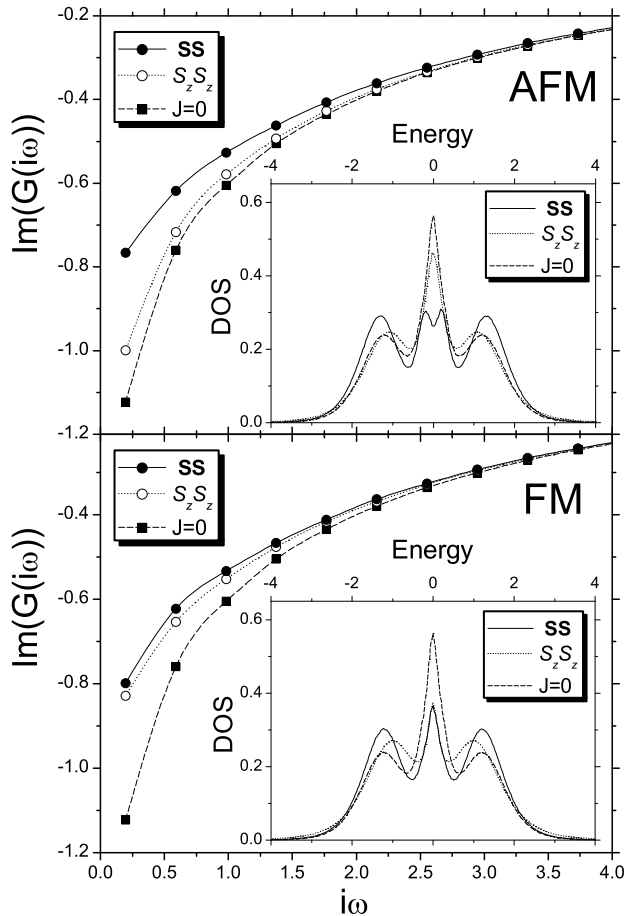}
\caption{\label{fig:epsart} Imaginary part of the Green functions
at Matsubara frequencies for the correlated adatom equilateral
triangle in the metallic bath for AFM (upper figure) and FM (lower
figure) types of effective exchange interaction. Parameters: $U=2,
J=\pm 0.2, t=0, \beta=16, \mu=U/2$. There are three dependencies
on each picture for $\mathbf{S}\mathbf{S}$, $S_zS_z$ and $J=0$
(which corresponds to single atom in the metallic bath) types of
interaction. The insets show DOS.}
\end{figure}

To explain these results we need to calculate the spectral density
$D(\omega)$ of the on-site spin-flip operators $S_{i}^{\pm}$ for
an AFM equilateral triangle
$D_{\mathbf{S}\mathbf{S}}(\omega)=\frac{2}{3}\delta(\omega)+
\frac{1}{3}\delta(\omega-3J)$, $
D_{S_zS_z}(\omega)=\frac{1}{3}\delta(\omega)+
\frac{2}{3}\delta(\omega-2J)$. In both cases there is a part of
the spectral density with zero frequency due to degeneracy of the
ground state and spin-flip transitions between its components, but
for the $\mathbf{S}\mathbf{S}$ case this part is twice as large.
For the case $J\gg T_K$ ($T_K$ is one-site Kondo temperature) only
this ``soft'' component of the spectral density will lead to Kondo
screening which means that the suppression of the Kondo effect is
twice more efficient for the $\mathbf{S}\mathbf{S}$ case than for
the $S_zS_z$ one. Since the trimer as a whole has a degenerate
ground state there is still a strong-coupling regime and an
effective Kondo temperature $T_{K}^{*}$ which however is much
smaller than $T_K$. Scaling considerations similar to one proposed
in Ref.\onlinecite{irkhin} gives an estimation of $T_{K}^{*}\simeq
T_{K}^{3/2}/J^{1/2}$ and $T_{K}^{*}\simeq T_{K}^{3}/J^2$ for the
$\mathbf{S}\mathbf{S}$ and $S_zS_z$ case respectively. We assume
that this quantity is too small to be visible in our simulations
(as well as in the experimental data \cite{CrAu}) so what is
observed corresponds to the one-site Kondo resonance at the
condition $T_{K}\geq J$. Similar estimation for the case of FM
interactions shows that there is no difference between the
$S_zS_z$ and $\mathbf{S}\mathbf{S}$ model there and
$T_{K}^{*}\simeq T_{K}^{3}/J^2$ as in the AFM $S_zS_z$ case.

In order to describe the experimental situation we changed the
geometry of the adatom trimer. An observation of the Kondo
resonance reconstruction was reported for one isosceles geometry
of three Cr atoms on a gold surface \cite{CrAu}. Thus we study
the isosceles triangles for AFM and FM types of effective exchange
interaction. We have chosen the following parameters of $J_{ij}$
to imitate the experimental system: $J_{23}=J, J_{12}=J/3,
J_{13}=J/3$. The computational results are presented in Fig. 3,
where one can see the reconstruction of resonance in AFM and FM
cases in accordance with experimental data. Note that the Kondo
resonance appears only for the more weakly bonded adatom in AFM case.

\begin{figure}
\includegraphics[width=\columnwidth]{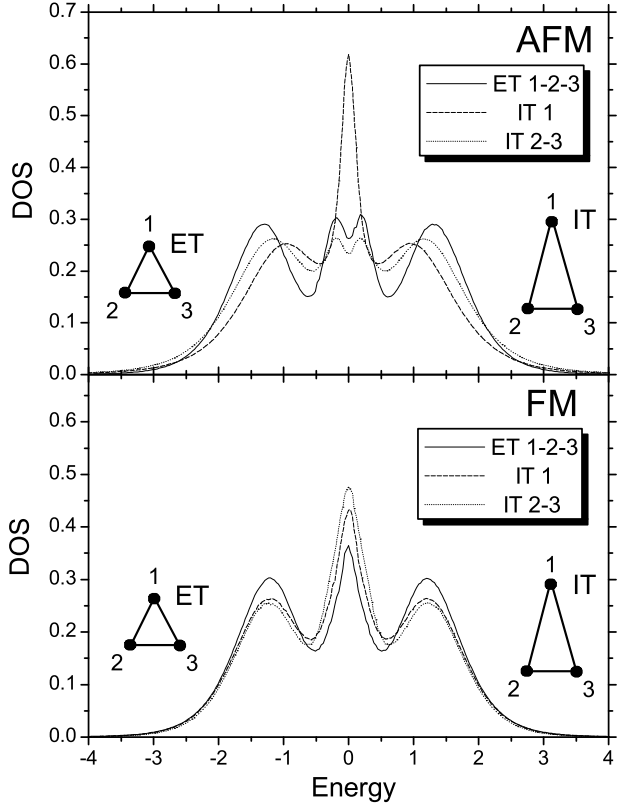}
\caption{\label{fig:epsart} DOS for equilateral triangle (ET) and
isosceles triangle (IT) geometries with AFM (upper figure) and FM
(lower figure) types of effective exchange interaction. Parameters
are the same as in Fig.2. Values of the effective exchange
integrals for IT are as follows: $J_{23}=J, J_{12}=J/3,
J_{13}=J/3$. There are two dependencies in case of IT: one for
adatom $1$ and another for equivalent adatoms $2$ and $3$. All
adatoms are equivalent in the case of ET (one dependence).}
\end{figure}

The observed picture can be drastically changed by introducing a
non-zero value of the hopping parameter $t_{ij}$. As noted above
the trimer ground state is degenerate at $t_{ij}=0$, however this
degeneracy is lifted for $t_{ij}\neq0$. If one of the obtained
states lies below the Fermi level, DOS can be changed drastically
and at certain parameters it leads to the appearance of a
resonance on the Fermi level (see Fig.4). In Fig.4, DOS is shown
for various values of filling in the system and at nonzero
$t_{ij}$. It is necessary to point out that the introduction of
the parameter $t_{ij}$, the variation of the filling in the system
($\mu$), and different geometries ($J_{ij}$) can lead to various
results in DOS. This can be used as one of the possible
explanations of the experimental data regarding the Cr trimer on a
Au surface which look initially appear counter-intuitive: the
Kondo temperature for the trimer is much larger than for the
single site, despite the suppression of the Kondo effect by
$J_{ij}$. One can assume that this is a consequence of the change
of the number of $d$-electrons in ground-state configuration for
the Cr atom in the trimer in comparison with the isolated one. In
order to describe specific experimental results quantitatively the
method developed here should be combined with first-principle
calculations in the spirit of recently developed ``local density
approximation plus dynamical mean-field theory'' (LDA + DMFT)
scheme \cite{ldadmft}.

\begin{figure}
\includegraphics[width=\columnwidth]{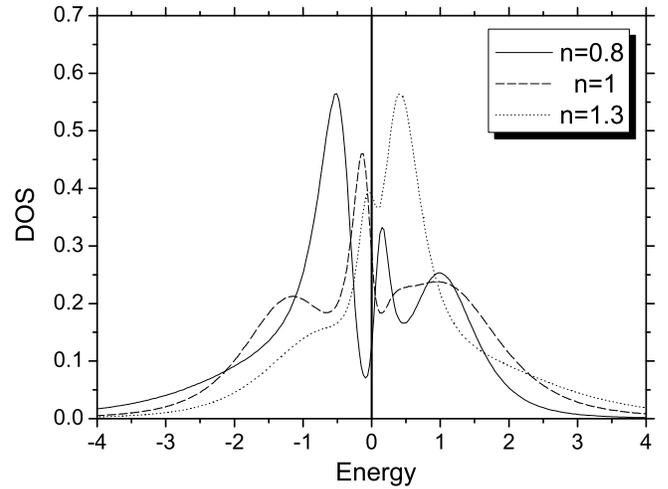}
\caption{\label{fig:epsart} DOS for equilateral triangle with an
AFM type of effective exchange interaction at various values of
filling in the system. Parameters: $U=2, J=0.2, t=0.3, \beta=16$.
Corresponding numbers of particles are $n=0.8$, $n=1$ (half-filled
case) and $n=1.3$.}
\end{figure}

In conclusion, we have shown that the electronic structure of a
correlated adatom trimer on a metallic surface drastically depends
on the symmetry of magnetic interactions. The effective exchange
interaction of $\mathbf{S}\mathbf{S}$ type leads to more efficient
suppression of the Kondo resonance in the AFM case than in the
case of $S_zS_z$ interactions. The experimental STM data
\cite{CrAu} can be reproduced qualitatively well by variation of
the geometry of the problem, hopping integrals and electronic
filling for magnetic nanosystems.

The work was supported by FOM project N0703M, NWO project
047.016.005 and "Dynasty" foundation.

\end{document}